\documentclass[twocolumn]{aastex631}

\usepackage{CJK}
\usepackage{amsmath}
\usepackage{amssymb}	
\usepackage{mathrsfs}

\newcommand{\msun}{M_\odot}

\definecolor{HHcolor}{rgb}{0.93,0.57,0.13}
\definecolor{HHcolor2}{rgb}{0.5,0.1,0.5}

\begin{document}
\begin{CJK*}{UTF8}{gbsn} 
\title{The Convergence of Heavy and Light Seeds to Overmassive Black Holes at Cosmic Dawn}
 
\correspondingauthor{Haojie Hu(胡豪杰)-JSPS Fellow}
\email{huhaojie@ccs.tsukuba.ac.jp}

\author[0000-0003-3143-3995]{Haojie Hu}
\affiliation{Center for Computational Science, University of Tsukuba, 1-1-1 Tennodai, 
Tsukuba, Ibaraki 305-8577, Japan}
\affiliation{Kavli Institute for Astronomy and Astrophysics, Peking University, 5 Yiheyuan 
Road,  Haidian District, Beijing, 100871, PRC}
\affiliation{Department of Astronomy, School of Physics, Peking University, 5 Yiheyuan Road,  
Haidian District, Beijing, 100871, PRC}

\author[0000-0001-9840-4959]{Kohei Inayoshi}
\affiliation{Kavli Institute for Astronomy and Astrophysics, Peking University, 5 Yiheyuan 
Road, Haidian District, Beijing, 100871, PRC}

\author[0000-0003-3633-5403]{Zolt\'an Haiman}
\affiliation{Department of Astronomy, Columbia University, New York, NY 10027, USA}
\affiliation{Department of Physics, Columbia University, New York, NY 10027, USA}
\affiliation{Institute of Science and Technoogy Austria, AM Campus 1, Klosterneuburg 3400 
Austria}

\author[0000-0001-6947-5846]{Luis C. Ho}
\affiliation{Kavli Institute for Astronomy and Astrophysics, Peking University, 5 Yiheyuan 
Road,  Haidian District, Beijing, 100871, PRC}
\affiliation{Department of Astronomy, School of Physics, Peking University, 5 Yiheyuan Road,  
Haidian District, Beijing, 100871, PRC}

\author{Ken Ohsuga}
\affiliation{Center for Computational Science, University of Tsukuba, 1-1-1 Tennodai, Tsukuba, 
Ibaraki 305-8577, Japan}


\begin{abstract}
The James Webb Space Telescope (JWST) has revealed low-luminosity active galactic nuclei (AGNs) 
at redshifts of $z\gtrsim 4-7$, many of which host accreting massive black holes (BHs) with 
BH-to-galaxy mass ($M_{\rm BH}/M_{\star}$) ratios exceeding the local values by more than an order of magnitude. 
The origin of these overmassive BHs remains unclear but requires potential contributions from 
heavy seeds and/or episodes of super-Eddington accretion. We present a growth model coupled with 
dark matter halo assembly to explore the evolution of the $M_{\rm BH}/M_{\star}$ ratio under 
different seeding and feedback scenarios. Given the gas inflow rates in protogalaxies, BHs 
grow episodically at moderate super-Eddington rates and the mass ratio increases early on, 
despite significant mass loss through feedback. Regardless of seeding mechanisms, the mass 
ratio converges to a universal value $\sim 0.1-0.3$, set by the balance between gas feeding 
and star formation efficiency in the nucleus. This behavior defines an attractor in the 
$M_{\rm BH}-M_{\star}$ diagram, where overmassive BHs grow more slowly than their hosts, while 
undermassive seeds experience rapid growth before aligning with the attractor. We derive an 
analytical expression for the universal mass ratio, linking it to feedback strength and halo growth.
The convergence of evolutionary tracks erases seeding information from the mass ratio by $z\sim 4-6$. 
Detecting BHs with $\sim 10^{5-6}~\msun$ at higher redshifts that deviate from convergence trend 
would provide key diagnostics of their birth conditions.
\end{abstract}

\keywords{Supermassive black holes (1663); Quasars (1319); High-redshift galaxies (734); Co-evolution}

\section{Introduction}\label{sec:intro}

The James Webb Space Telescope (JWST) with its exceptional sensitivity has enabled the discovery of 
faint active galactic nuclei (AGNs) in the high-redshift universe 
\citep[$z > 4-7$, e.g.,][]{Onoue_2023,Kocevski_2024}, pushing the detection limits of AGN luminosities 
to $L_{\rm bol} \sim 10^{45}~{\rm erg~s}^{-1}$ and black hole (BH) masses to 
$M_{\rm BH} \sim 10^6~M_\odot$ \citep[e.g.,][]{Maiolino_2023,Harikane_2023}. A significant fraction 
of these newly discovered AGNs host supermassive BHs (SMBHs) with BH-to-host galaxy mass ratios 
reaching values of $M_{\rm BH}/M_\star\gtrsim 0.01-0.1$ \citep[][]{Chen_2024}, significantly higher 
than the empirical value ($\lesssim 0.01$) observed in the present-day universe by more than an 
order of magnitude \citep[e.g.,][]{Kormendy_2013,Reines_2015}. While such overmassive BHs were previously 
identified in rare, ultra-luminous quasars (\citealt[][]{Mortlock_2011,Wu_2015,Banados_2018,Fan_2023}, 
also local overmassive SMBHs, e.g., \citealt[][]{van_2012}), JWST has now revealed similar trends in fainter, 
more representative AGN populations, providing strong constraints on the early stages of BH-galaxy 
co-evolution as well as the assembly process of each component \citep[e.g.,][]{Trinca_2022, Li_2024, Li_2023}.

The rapid assembly of SMBHs in the first billion years of the universe have been extensively debated
\citep[e.g.,][and references therein]{Inayoshi_2020,Volonteri_2021}. Two main scenarios have been proposed: 
(1) rapid gas accretion of stellar remnant BHs at or above the Eddington rate with a high duty cycle 
from early epochs at $z \gtrsim 20$ \citep{Madau_2001, Haiman_2001,Volonteri_2003}, and
(2) moderate growth from more massive seed BHs formed via direct collapse of massive primordial gas 
clouds or runaway stellar collisions in dense environments 
\citep[e.g.,][]{Bromm_2003,Regan_2009a, Regan_2009b,Li_2023a}. These mechanisms are widely considered 
essential for explaining the high BH-to-stellar mass ratio observed in JWST-identified AGNs 
\citep[e.g.,][]{Inayoshi_2022a,Hu_2022c,Scoggins_2024}. While some extreme cases may originate from 
massive seed BHs, which naturally yield high BH-to-galaxy mass ratios at birth 
\citep[e.g., an AGN candidate with $M_{\rm BH}/M_\star \simeq 1.0$;][]{Bogdan_2024}, intermittent 
super-Eddington accretion could also drive even light seed BHs to outgrow their hosts if sustained 
gas inflows are available \citep[][]{Inayoshi_2022a,Hu_2022c,Scoggins_2024}. Despite both pathways 
being plausible, a key challenge is understanding how BH-host interactions shape their long-term 
evolutionary trajectories.

AGN and stellar feedback play a central role in regulating BH growth. However, most cosmological 
galaxy-formation simulations, which rely on subgrid feedback models due to limited spatial resolution, 
predict substantially lower $M_{\rm BH}/M_\star$ ratios than those observed, i.e., undermassive BHs.
According to these models, galaxies grow first through star formation, while BH growth is delayed 
until galaxies become massive enough to retain gas in the nuclei due to their deeper gravitational 
potential wells \citep[][]{Dubois_2015,Habouzit_2017,Angles-Alcazar_2017,Habouzit_2025}. 
This tension between these simulation results and the JWST-identified AGNs/BHs with high 
$M_{\rm BH}/M_\star$ ratios highlights the need for a revised treatment of feedback processes in 
galactic nuclei at early times.

In this Letter, we present a BH growth model coupled with DM halo assembly to explore the emergence of 
overmassive BH populations with $M_{\rm BH}/M_\star>0.01$, exceeding the values observed in the local 
universe. We parameterize feedback-driven mass loss by modeling gas inflow rates toward the nucleus 
as a power-law function, $\dot{M}(r)\propto r^p$, where $p$ encapsulates uncertainties of feedback effects.
By incorporating BH and galaxy growth rates tied to DM halo evolution, we develop a semi-analytical 
framework to map evolutionary trajectories in the $M_{\rm BH}-M_\star$ diagram. The model successfully 
reproduces the $M_{\rm BH}-M_\star$ distribution of JWST-identified AGNs and BHs at $z\sim 4-7$ 
\citep[e.g.,][]{Kokorev_2023,Bogdan_2024,Greene_2024} across different seeding channels, even with 
significant mass loss through feedback. We further discuss how rapid assembly processes diminish 
seeding information from $M_{\rm BH}/M_\star$ ratios of observed AGN populations. Throughout this paper, 
we assume a flat $\Lambda$ cold dark-matter universe with the following cosmological parameters: 
$h=0.6732$, $\Omega_{\rm m}=0.3158$, and $\sigma_8=0.8120$ \citep{Planck_2020}.

\vspace{3mm}
\section{The Early Growth of Seed BH and Evolution of $M_{\rm BH}/M_{\star}$ Ratio} \label{sec:cosmic}

In this study, we construct an analytical framework to model the mass growth of BHs and their host galaxies.
In the hierarchical structure formation paradigm, we link the rate of baryon inflow into the galactic 
nucleus to the DM halo mass assembly as
\begin{equation}
\dot{M}_0 = \epsilon_{\rm nuc}f_{\rm b}~\frac{{\rm d} M_{\rm halo}(t)}{{\rm d} t},
\label{eq:a}
\end{equation}
where $f_{\rm b}=0.16$ is the baryonic fraction \citep{Planck_2020}, and ${\rm d}M_{\rm halo}/{\rm d}t$ is 
the DM halo growth rate
obtained through a Monte Carlo merger tree algorithm \citep[][]{Parkinson_2008,Li_2021}.
Generally, incoming gas from the parent halo is assembled into a galactic disk, and then consumed 
either by star formation or BH accretion, or possibly ejected into galactic outflows 
\citep[][]{Wadsley_2004,Fraternali_2005,Kaufmann_2006,Fraternali_2006}.
We assume that a fraction $\epsilon_{\rm nuc}$ of the gas inflow feeds the nuclear region, 
and set a fiducial value of $\epsilon_{\rm nuc}=0.1$ based on the results obtained by cosmological 
hydrodynamic simulations resolving multiscales down to $\sim 0.1$ pc \citep{Hopkins_2010}.

\begin{figure*}[t!]
\centering
\includegraphics[scale=0.49]{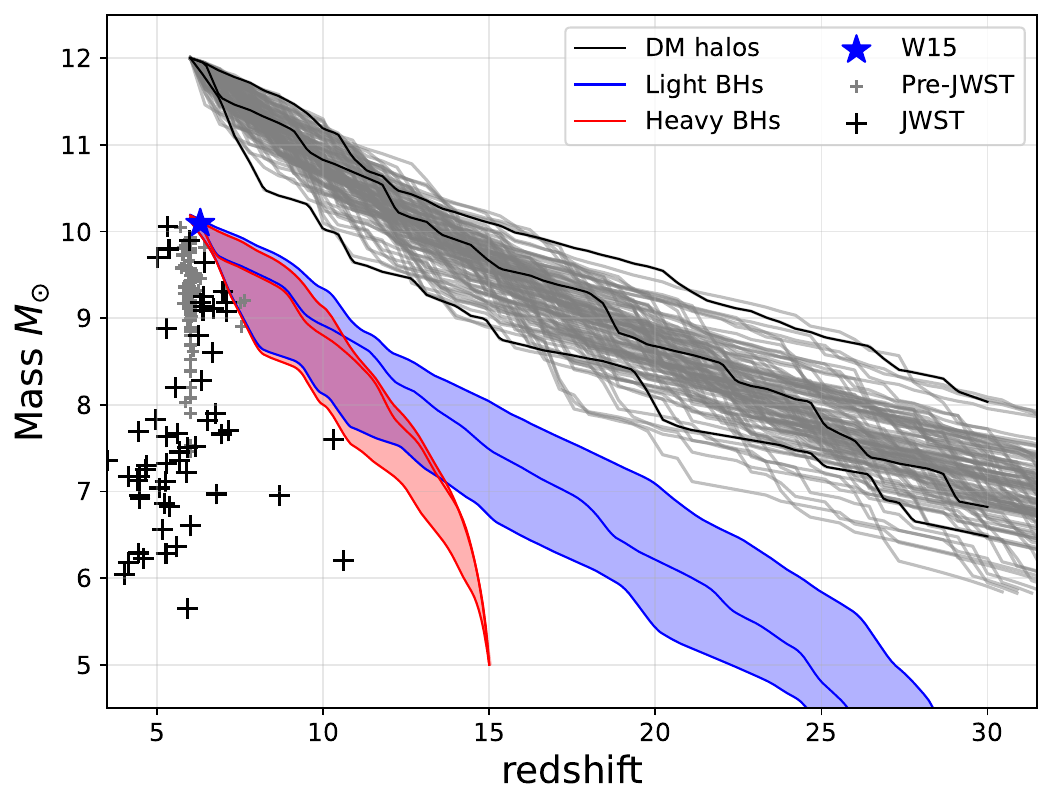}
\includegraphics[scale=0.49]{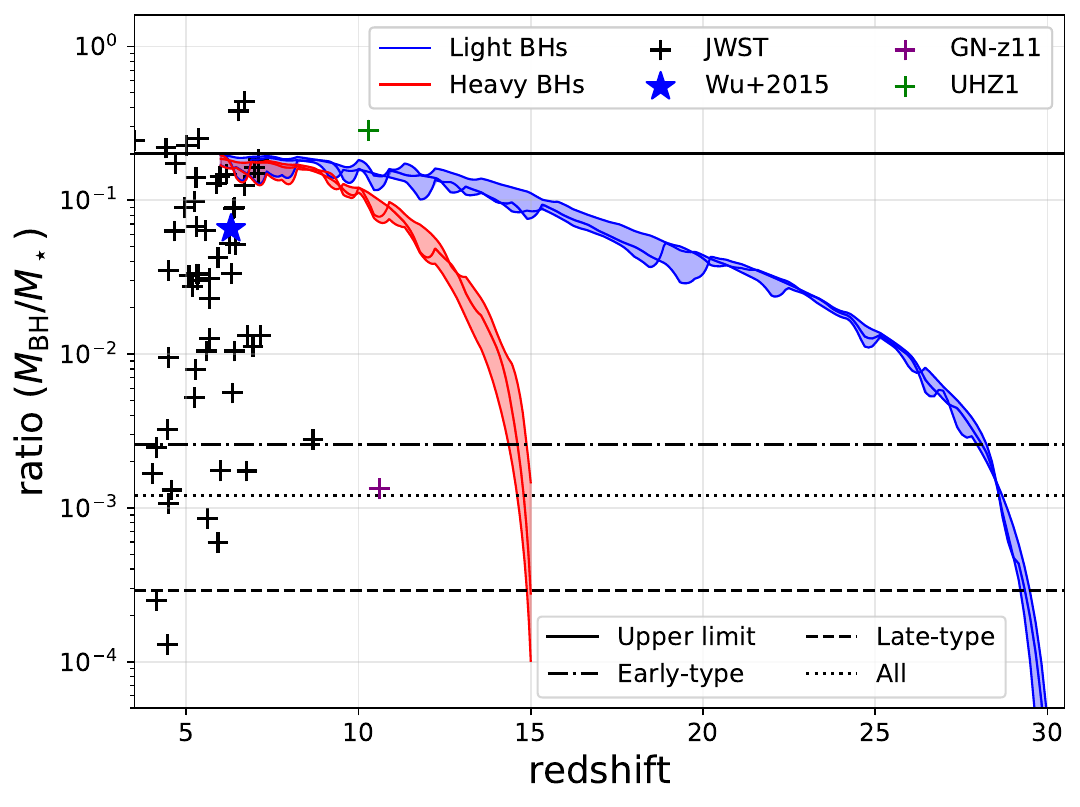}
\caption{
Left: The mass evolution for DM mergers and seed BHs. The grey curves are evolutionary tracks 
of main progenitors of $100$ (out of $10^4$) DM halos ended up as massive halo with 
$M_{\rm h}=10^{12}~M_\odot$ at $z=6$. Three typical merger histories are shown in black solid 
curves. The growth rate $(k_{\rm h})$ for different merger trees at $z\sim 6$ varies from 
$\sim 0.25$ to $\sim 1.15$. The blue and red curves are evolutionary tracks for 
light ($10~M_{\odot}$ BH at $z = 30$) and heavy seed BHs ($10^5~M_\odot$ at $z = 15$) 
based on the three merger trees, adopting a fiducial model with $p=0.5$. 
The blue and red shaded regions are parameters for possible SMBHs from our models.
The high redshift SMBHs are overlaid in crosses, with the most massive SMBH from \citet[][]{Wu_2015}.
SMBHs from pre-JWST era are overlaid in grey crosses \citep[][]{Mortlock_2011,Banados_2018,Izumi_2019},
while SMBHs from JWST era are shown in black crosses 
\citep[data collected from][]{Ding_2023,Maiolino_2023,Harikane_2023,Kocevski_2023,Larson_2023,Yue_2023,
Kokorev_2023,Stone_2024,Maiolino_2024,Juodzbalis_2024,Bogdan_2024,Greene_2024,Kocevski_2024}.
Right: The evolution of the $M_{\rm BH}/M_{\star}$ ratio for light seed BHs (blue curves) 
and heavy seed BHs (red curves). The JWST samples are overlaid for reference. The solid horizontal line 
indicates the maximum value of $M_{\rm BH}/M_{\star}\sim {\epsilon_{\rm nuc}}/{\epsilon_\star}=0.2$. 
The dashed, dash-dotted and dotted horizontal lines are $M_{\rm BH}/M_{\star}$ ratios for late-type, 
early-type and all galaxies at $M_\star=3\times 10^{10}~\msun$, respectively \citep[][]{Greene_2020}.
\label{fig:1}}
\vspace{5mm}
\end{figure*}

We assume the BH accretion rate to match the nuclear gas supply rate 
i.e., $\dot{M}_{\rm BH} = \dot{M}_0$, when it is below the Eddington limit.
In contrast, when the inflow rate exceeds this limit, we consider mass removal 
via radiation-driven outflows to reduce the BH feeding rate as \citep[][]{Hu_2022c},
\begin{equation}
\dot{M}_{\rm BH} = \dot{M}_0\times {\rm min}\left\{1, \left(\frac{r_{\rm in}}{r_{\rm out}}\right)^p \right\}, 
\label{eq:a}
\end{equation}
where $r_{\rm in}$ and $r_{\rm out}$ are the inner and outer radii, where the gas inflow rate 
decreases toward the center. The power-law index of $p$ characterizes the outflow strength and 
the efficiency of mass removal by outflows. The index is measured as $p\simeq 0.5$ for 
super-Eddington accretion cases \citep{Hu_2022c}, but it ranges over $0<p<1$ in more general 
situations \citep[see][]{Yuan_2014}. In this work, we set $r_{\rm in}$ to the innermost stable 
circular orbit (ISCO) radius, and $r_{\rm out}$ to the photon-trapping radius in the accretion 
flow, i.e., $r_{\rm in }=6GM_{\rm BH}/c^2$ and $r_{\rm out}=\kappa_{\rm T} \dot{M}_0/(4\pi c)$,
with $G$ the gravitational constant, $c$ the speed of light and $\kappa_{\rm T}$ the electron 
scattering opacity. The transition due to outflows occurs when 
$\dot{m}_0(\equiv \dot{M}_0/\dot{M}_{\rm Edd})$ exceeds 0.6, where the Eddington accretion rate 
is defined as $\dot{M}_{\rm Edd}\equiv L_{\rm Edd}/(0.1c^2)$ and $L_{\rm Edd}$ is the Eddington luminosity.
As a result, the BH accretion rate in Equation (\ref{eq:a}) is rewritten as
\begin{equation}
\dot{M}_{\rm BH}= \dot{M}_0\times {\rm min}\left\{ 1, \left( \frac{\dot{m}_0}{0.6}\right)^{-p}\right\}.
\label{eq:d}
\end{equation}

To quantify the stellar mass evolution in a galaxy, we assume a star formation efficiency (SFE) 
$\epsilon_\star$, where the star formation rate is given by 
$\dot{M}_\star = \epsilon_\star \dot{M}_0 / \epsilon_{\rm nuc}$. We here only consider a high 
and constant SFE model ($\epsilon_\star=0.5$), which is motivated by JWST observations 
of UV bright galaxies at $z>10$, whose SFEs are expected to be as high as $\sim 0.3-0.5$ 
\citep[e.g.,][]{Wang_2023,Inayoshi_2022b,Pontoppidan_2022}. This model provides a higher stellar 
mass for a given inflow rate and thus lower $M_{\rm BH}/M_\star$ ratios, compared to those assumed 
in previous semi-analytical studies \citep[e.g.,][]{Behroozi_2019,Scoggins_2024}.

In the left panel of Figure~\ref{fig:1}, we present evolutionary tracks for DM halo mergers and BHs, 
along with distribution of high-redshift quasars observed with JWST \citep[black crosses,][]{Ding_2023,Maiolino_2023,Harikane_2023,Kocevski_2023,Larson_2023,Yue_2023,Kokorev_2023,Stone_2024,
Maiolino_2024,Juodzbalis_2024,Bogdan_2024,Greene_2024,Kocevski_2024} 
and with other surveys \citep[gray crosses,][]{Mortlock_2011,Wu_2015,Banados_2018,Izumi_2019}. The 
gray solid curves are evolutionary tracks of the main DM halo progenitors that end up in massive 
halos with $M_{\rm h} = 10^{12}~M_\odot$ at $z = 6$ \citep[][]{Li_2021}, the characteristic halo 
predicted by Press-Schechter formalism, when comparing the growth rate of quasar density to 
observations \citep[][]{Wyithe_2006}. As examples, we highlight three representative 
merger trees (black curves) and the corresponding mass evolution of BHs originating from two seeding 
scenarios BH that are initiated and grow within these representative halos. The light seed BHs (blue 
curves) are originated from Pop III stellar remnants at $z=30$ with $M_{\rm BH}=10~M_\odot$, while the 
heavy seed BHs (red curves) begin their mass growth at $z=15$ with initial masses of 
$M_{\rm BH}=10^5~M_\odot$. Both types of seed BH undergo early rapid growth phases, and all tracks reach 
$M_{\rm BH} \simeq 10^{10}~M_\odot$ by $z \sim 6$. The BH mass and redshift are consistent with those 
of the brightest quasars at $z\sim 6$ \citep{Wu_2015}.

In the right panel of Figure~\ref{fig:1}, we show the evolution of the $M_{\rm BH}/M_\star$ ratio
as a function of redshift for the two BH seeding scenarios in the three DM halo trees. In both cases,
the ratio increases rapidly during the early stage and overshoots the empirical value observed in the 
local universe. Although the galaxy gains stellar mass efficiently with a high SFE ($\epsilon_\star =0.5$), 
the rapid growth of BHs drives the ratio to $M_{\rm BH}/M_\star \sim 0.1$ by $z \sim 6$, consistent with
observations of overmassive SMBHs \citep[e.g.,][]{Harikane_2023,Maiolino_2023,Bogdan_2024}.
It is worth noting that the final mass ratio is determined by the last accretion episodes at $z\sim 6$, 
when the growth rate transitions to a sub-Eddington value. In the sub-Eddington regime, all inflowing 
gas is assumed to feed the central BH without significant mass loss, naturally leading to 
$M_{\rm BH}/M_\star \simeq {\epsilon_{\rm nuc}}/{\epsilon_\star}(=0.2)$. We note that the mass 
ratio might be reduced if the AGN feedback operates even in the sub-Eddington regime, where we 
currently set $p=0$ in this model.

In this framework, the rapid growth phases of seed BHs make the distinction between seeding scenarios 
challenging. Both light and heavy seed BHs converge in their evolutionary tracks by $z \lesssim 10$, 
making the $M_{\rm BH}/M_\star$ ratio an inefficient discriminator of seeding models. Therefore, 
observing the early evolutionary stages of less massive BHs may help break the degeneracy.

\section{Global flow structure on the $M_{\rm BH}-M_\star$ diagram.}
\label{sec:flowpatterns}

\subsection{Analytical formulation}
\label{subsec:analytical}

In the previous section, we developed evolutionary tracks for different seed BHs along specific halo 
assembly histories and investigated the evolution of the $M_{\rm BH}/M_\star$ ratio to account for the 
presence of luminous quasars at $z \sim 6$. In this section, we extend our analysis to explore the 
broader evolutionary trends of $M_{\rm BH}/M_\star$ ratios across a wider range of parameters. 
Specifically, we examine the patterns in the $M_{\rm BH}-M_\star$ diagram for the entire BH population, 
including the less luminous and less massive BHs recently uncovered by JWST observations 
\citep[e.g.,][]{Kocevski_2023, Harikane_2023, Maiolino_2023}.

To achieve this, we generalize the halo assembly tracks and extend the model to cover a broader 
parameter space. This allows us to explore the $M_{\rm BH}/M_\star$ ratio for SMBHs of varying 
masses in different galaxies, using the following analytical formula
\begin{equation}
M_{\rm halo}(z) = \mathcal{M} \cdot \exp (-k_{\rm h}z),
\label{eq:e}
\end{equation}
where $\mathcal{M}$ is the mass normalization and $k_{\rm h}$ represents the growth rate of the DM halo. 
This redshift dependence arises from the Press-Schechter formalism \citep[][]{Press_1974,Lacey_1993,Cole_2000} 
and aligns with fits to merger trees from cosmological $N$-body simulations \citep{Fakhouri_2010,Dekel_2013},
yielding the mass growth rate of $d\ln M_{\rm h}/dt \propto k_{\rm h}(1+z)^{5/2}$. The mean value of 
$k_{\rm h}$ for all merger trees is estimated from comparison to these simulations 
\citep[e.g.,][]{Fakhouri_2010} as $\langle k_{\rm h}\rangle \simeq 0.7$. For halos reaching 
$M_{\rm halo}= 10^{12}~M_\odot$ at $z=6$, the power-law index ranges over 
$0.25\lesssim k_{\rm h} \lesssim 1.15$, as illustrated by the three representative merger trees in 
Figure~\ref{fig:1}.

Using this functional form for stellar mass growth with an SFE of $\epsilon_\star$, the ratio of 
$\dot{M}_{\rm BH}/\dot{M}_\star$ is calculated as
\begin{equation}
\frac{\dot{M}_{\rm BH}}{\dot{M}_\star}=
\frac{{\rm d}M_{\rm BH}}{{\rm d}M_\star}=
C_p(z)\left(\frac{M_{\rm BH}}{M_\star}\right)^p,
\label{eq:Mdot_ratio}
\end{equation}
and
\begin{equation}
C_p(z)= \left(\frac{t_{\rm Edd}}{t_\star}\right)^{{-p}} 
\left(\frac{\epsilon_{\rm nuc}}{\epsilon_\star}\right)^{1-p},
\label{eq:Cp}
\end{equation}
where $t_{\rm Edd}\simeq 45~{\rm Myr}$ is the $e$-folding timescale for the Eddington-limited growth with 
a 10\% radiative efficiency, and the characteristic star formation timescale $t_\star$ is defined as 
\begin{equation}
t_\star = \frac{3}{5H_0(1+z)E(z)k_{\rm h}}\simeq 49~{\rm Myr}~k_{\rm h}^{-1}\left(\frac{1+z}{10}\right)^{-5/2},
\label{eq:e}
\end{equation}
where $E(z)=[(1+z)^3 \Omega_{\rm m}+\Omega_\Lambda]^{1/2}$
and $z>(\Omega_\Lambda/\Omega_{\rm m})^{1/3}-1\simeq 0.3$ 
is considered. 
Therefore, for given values of initial $M_{\rm BH}/M_\star$, $p$, $k_{\rm h}$, and $z$, one can 
numerically calculate the ratio of the growth rates of the BH and stellar masses and draw the 
``velocity'' vectors on the $M_\star-M_{\rm BH}$ diagram.

\begin{figure*}[t!]
\centering
\includegraphics[scale=0.58]{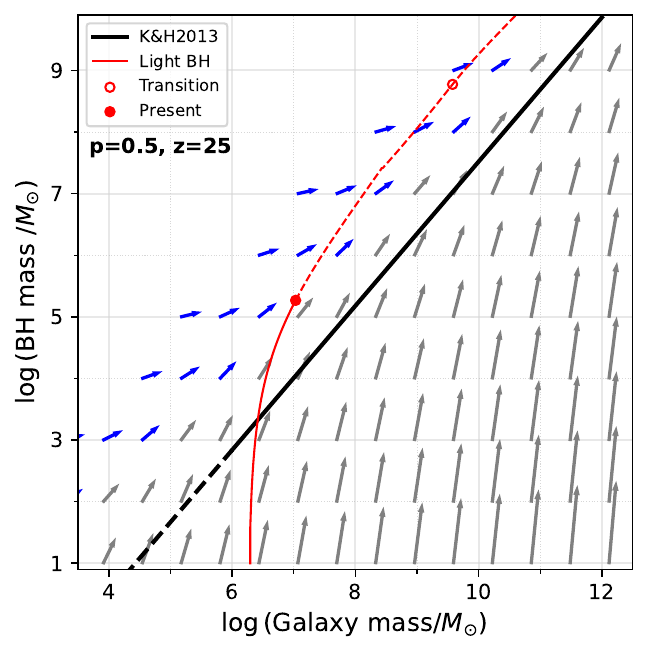}\hspace{2mm}
\includegraphics[scale=0.58]{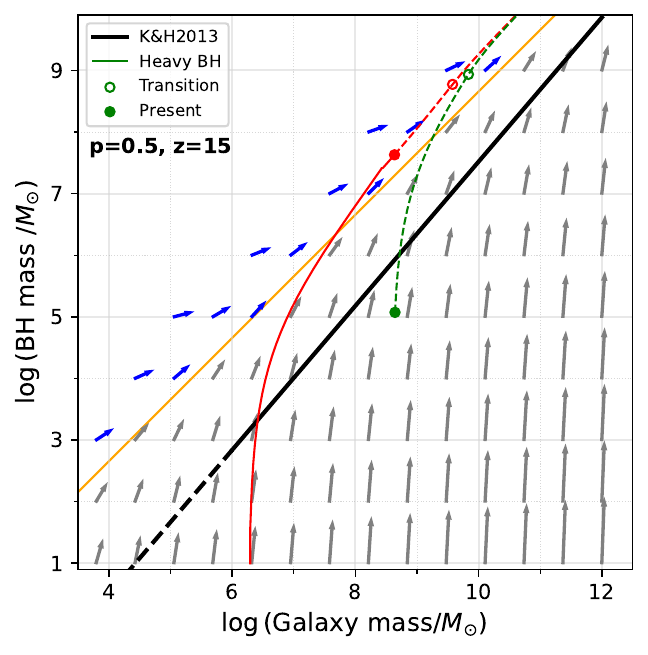}\\\vspace{2mm}
\includegraphics[scale=0.58]{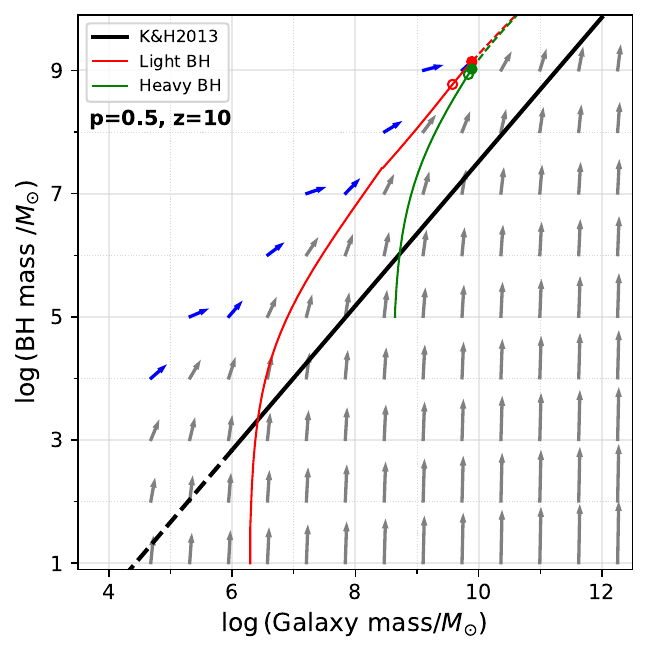}\hspace{2mm}
\includegraphics[scale=0.58]{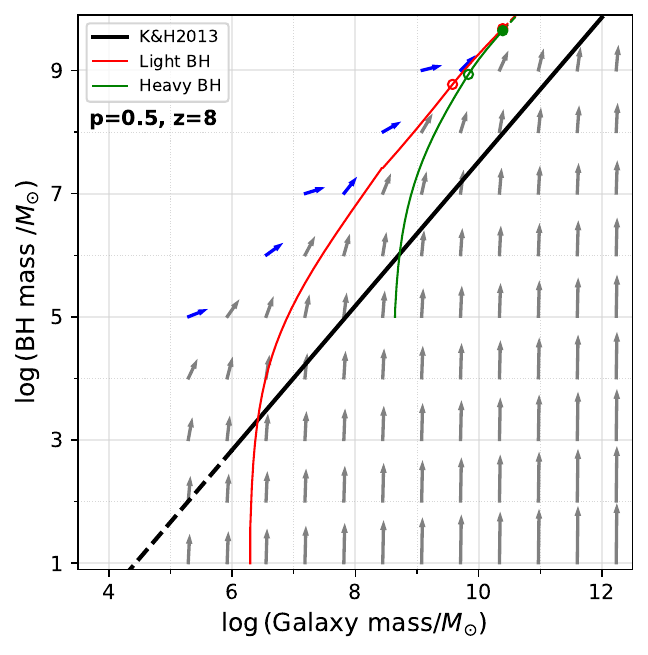}
\caption{
The evolutionary trend for two different seed BHs at four epochs ($z=25$, $15$, $10$, and $8$).
The blue (gray) arrows indicate positive (negative) 2nd-order derivatives $\mathscr{S}$, while lengths of arrows are 
calculated assuming a $10~$Myr growth for BH and host galaxy at the current epoch and renormalized 
to $[0.2,~1.2]$ times the length of a unit vector. 
The two evolutionary tracks represent that for a light seed with an initial mass of $M_{\rm BH}=10~\msun$ at $z=30$ (red) and 
a heavy seed with $M_{\rm BH}=10^5~\msun$ at $z=15$ (green).
The solid curve illustrates the evolution from the seeding time to the redshift of each panel (marked with filled circles), 
while the dashed curves show the continuation of the track toward lower redshifts.
In this figure, we adopt a constant value $p=0.5$ and halo growth rates $k_{\rm h}=0.7$ at $z\leq 15$ and $k_{\rm h}=0.35$ at $z>15$,
to be consistent with the evolutionary tracks shown in Figure~\ref{fig:1}.
The open circle in each model indicates the transition from super-Eddington to sub-Eddington accretion, after which $p=0$ is assumed.
The diagonal orange line in the top-right panel represents the boundary with $\mathscr{S}=0$ for $p=0.5$.
\label{fig:2}}
\vspace{3mm}
\end{figure*}

\begin{figure*}[t!]
\centering
\includegraphics[scale=0.51]{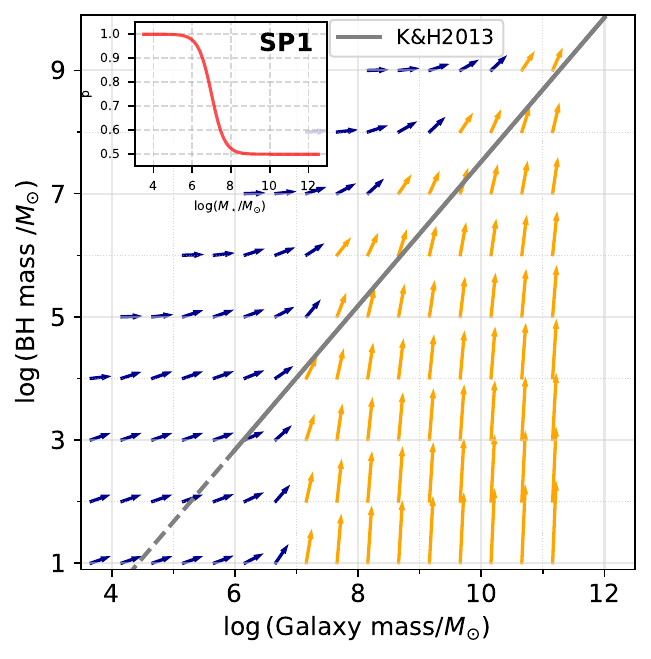}
\includegraphics[scale=0.51]{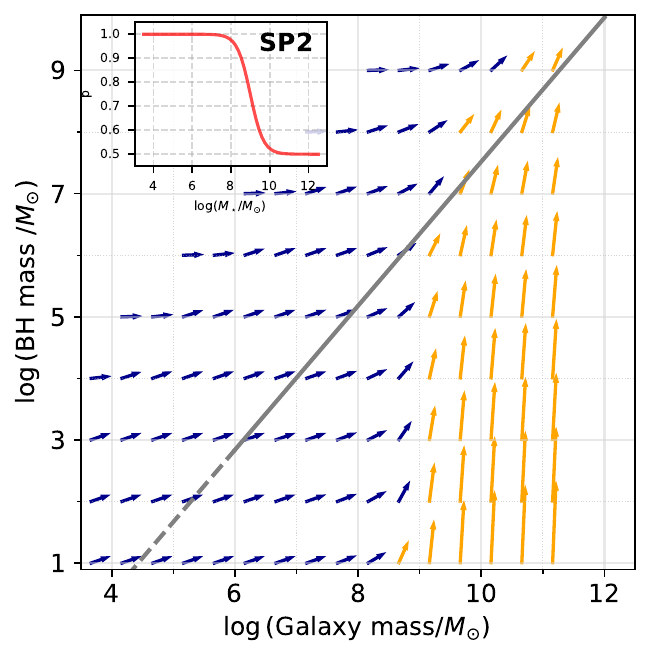}
\includegraphics[scale=0.51]{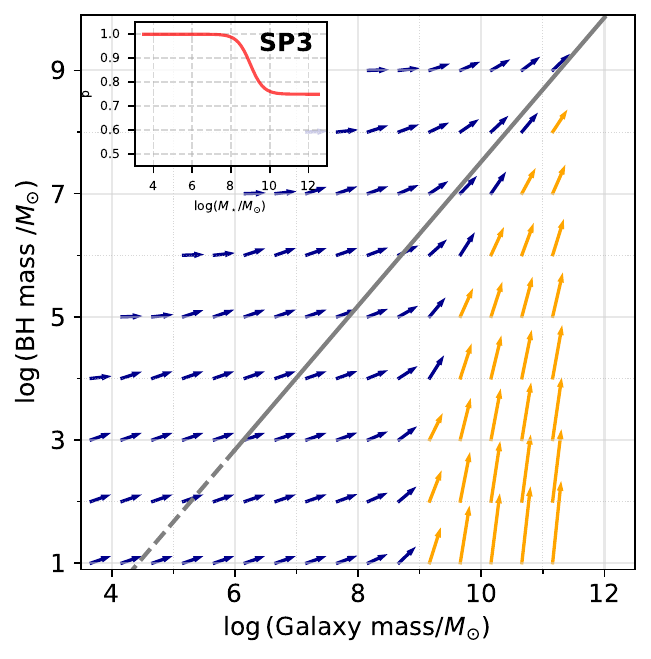}
\caption{
The evolutionary trend for different feedback models at $z=15$ (from left to right: 
SP1, SP2, SP3). The blue arrows indicate the positive signs of second-order derivatives, while the orange 
arrows are for negative ones. The lengths of arrows are same as in Figure~\ref{fig:2}. The inserts in 
each panel show the corresponding feedback models.
\label{fig:3}}
\end{figure*}

The behavior of the evolutionary track can be understood in a simple way because the numerical factor $C_p(z)$ 
evolves slowly with redshift compared to the stellar mass growth, as described by
\begin{equation}
\frac{{\rm d}\ln C_p}{{\rm d}\ln M_\star} = \frac{5p/2}{k_{\rm h}(1+z)}\simeq 0.16,
\end{equation}
for $p=0.5$, $k_{\rm h}=0.7$, and $z=10$.
Thus, when $C_p$ is approximated as a constant value (due to a small $p$-value or high redshift), 
the mass ratio $M_{\rm BH}/M_\star$ asymptotically approaches
\begin{equation}
\frac{M_{\rm BH}}{M_\star}~\rightarrow ~C_p^{\frac{1}{1-p}}
\label{eq:mass_ratio}
\end{equation}
regardless of the $p$-value in the limit of weak or moderate feedback ($p\lesssim 0.5$).
In this regime, the logarithmic gradient approaches unity as
\begin{equation}
\frac{{\rm d}\ln M_{\rm BH}}{{\rm d} \ln M_\star} = C_{ p}\left(\frac{M_{\rm BH}}{M_\star}\right)^{p-1}\rightarrow 1.
\label{eq:mass_ratio_gradient}
\end{equation}
This asymptotic behavior explains the flow pattern and convergence shown in Figure~\ref{fig:2}.

Figure~\ref{fig:2} illustrates the evolutionary trend of the BH-to-stellar mass ratio at four different 
redshifts: $z=25$, $15$, $10$, and $8$. As a reference, the local relation from \citet{Kormendy_2013} 
is overlaid with the black diagonal line. We adopt a constant value $p=0.5$ and halo growth rates 
$k_{\rm h}=0.7$ at $z\leq 15$ and $k_{\rm h}=0.35$ at $z>15$, to be consistent with the evolutionary 
tracks shown in Figure~\ref{fig:1}. The arrows indicate the evolutionary direction for given boundary 
conditions, i.e., BH mass, halo growth, and feedback strength. Their lengths are calculated based on 
the growth of BH mass and stellar masses over the next $10$ Myr and re-normalized to $[0.2,~1.2]$ times 
the length of a unit vector for visualization. The two-dimensional plane is divided into two regions 
based on the gradient of the vector field (i.e., acceleration in the analogy of fluid dynamics), 
$\mathscr{S}\equiv d^2\ln M_{\rm BH}/(d\ln M_\star)^2$: the region with blue arrows where $\mathscr{S}>0$ 
and region with gray arrows where $\mathscr{S}<0$, respectively. The boundary where $\mathscr{S}=0$ is shown 
as a diagonal orange line in the $M_{\rm BH}-M_\star$ diagram, as derived in 
Equation~(\ref{eq:mass_ratio_gradient}). Below the boundary of $\mathscr{S}=0$, the BH mass tends to grow 
faster than the stellar mass and approaches the boundary. On the other hand, above the boundary, the 
stellar mass grows faster than the BH mass and approaches the boundary. In both cases, the boundary of 
$\mathscr{S}=0$ becomes an attractor and the mass ratio approaches a constant value given by 
Equation~(\ref{eq:mass_ratio}).

As examples, we overlay the evolutionary tracks of two types of seed BHs: a light seed BH with an 
initial mass of $M_{\rm BH}=10~\msun$ at $z=30$ and a heavy seed BH with $M_{\rm BH}=10^{5}~\msun$ 
at $z=15$. Each track features an open circle representing the transition from super-Eddington 
accretion regime to sub-Eddington regime. The solid curves illustrate the evolution from the 
seeding phase to that redshift, while the dashed curves indicate the continuation of the tracks 
toward lower redshifts. Initially, the two cases have BHs undermassive relative to their host 
galaxies and lie below the local relation \citep[][]{Kormendy_2013} with $\mathscr{S}<0$. 
Given the fiducial feedback strength ($p=0.5$), these seed BHs experience rapid but decelerating 
growth during their early evolutionary stages. This initial rapid growth increases the 
${M_{\rm BH}}/{M_\star}$ ratio to values significantly above the local relationship. After that, 
the two evolutionary tracks nearly converge to a similar locus by $z\simeq 6$, making it
difficult to distinguish between the BH seeding models based on the observed mass ratio at $z\lesssim 6$.
The converging mass ratio is primarily determined by the feedback strength $p$ (see also Equation~\ref{eq:mass_ratio}).
However, as the gas-supply rate to the galactic nucleus transitions to the sub-Eddington regime 
(evolutionary tracks after open circles, $p=0$ is assumed in the feedback prescription) at lower redshifts, 
the final mass ratio approaches $M_{\rm BH}/M_\star= \epsilon_{\rm nuc}/\epsilon_\star (\simeq 0.2)$.

\begin{figure*}
\centering
\includegraphics[scale=0.60]{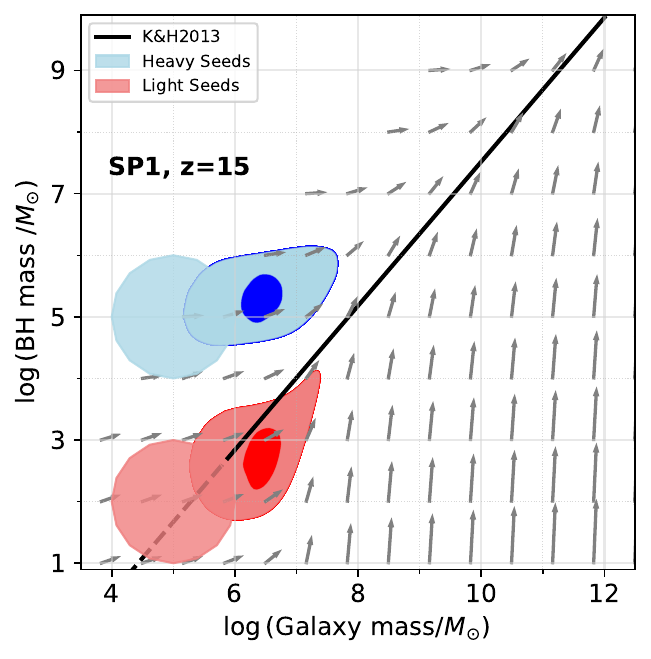}
\includegraphics[scale=0.60]{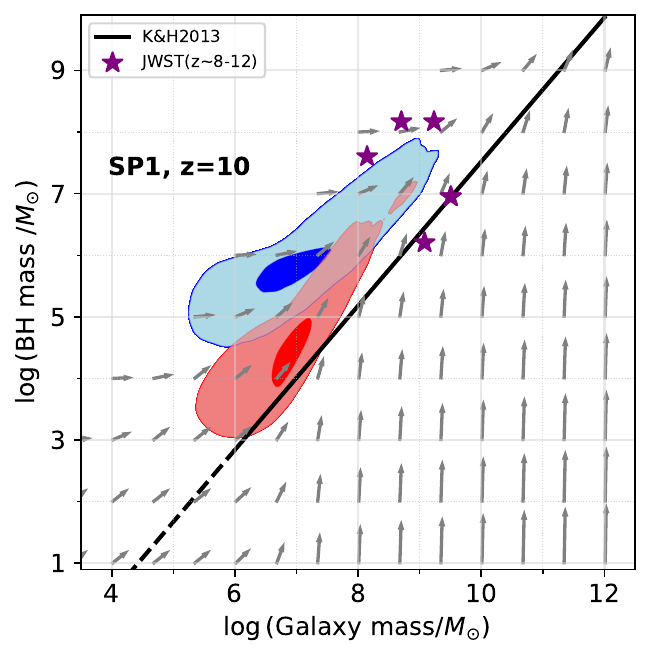}\\
\includegraphics[scale=0.60]{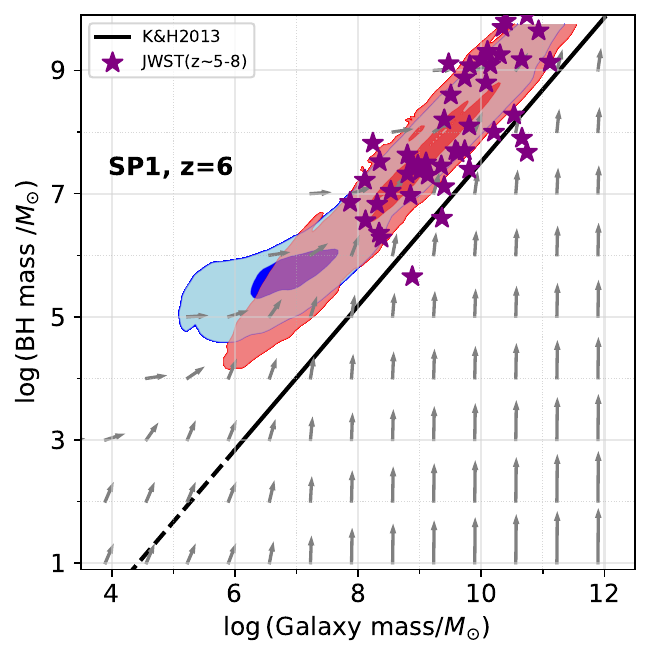}
\includegraphics[scale=0.60]{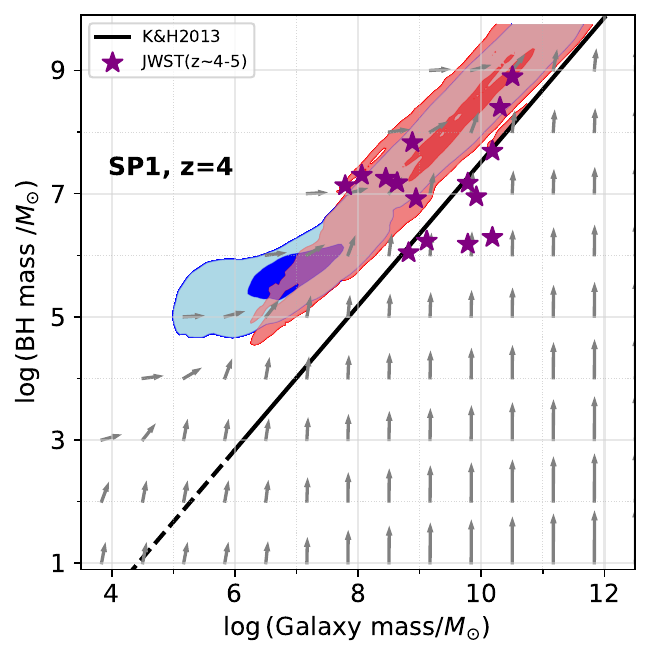}
\caption{
The evolutionary tracks for two classes of seed BHs harbored in galaxies on different mass scales: 
the blue region for heavy seed BHs; the red region for light seed BHs. The flow patterns are calculated 
at each redshift adopting the SP1 feedback model. The seed BHs are chosen so that they are located above, 
and below the local relation when they were seeded at $z=15$. The BH-galaxy systems evolve to $z=4$ 
with JWST SMBHs overlaid as stars in each redshift bin. For our fiducial parameters, we adopt 
$k_{\rm h}=0.7$ and $\epsilon_{\rm nuc}=0.02$. The two colored contours are the probability distributions
of the evolutionary tracks for the two seed BH models when normal distributions for $k_{\rm h}$ and 
$\epsilon_{\rm nuc}$ are adopted, with $\langle k_{\rm h}\rangle=0.70$, $\sigma_{k}^2=0.06$, and 
$\langle \log\epsilon_{\rm nuc}\rangle =-1.7$, $\sigma_{\rm nuc}^2=0.25$. The contour lines represent 
coverages of $16\%$, and $84\%$, respectively.
}
\label{fig:4}
\end{figure*}

\subsection{The variation due to feedback models}\label{sec:evolution}

While radiation hydrodynamic simulations that focus on accretion flows at the vicinity of a BH 
event horizon suggest that the power-law index of the inflow rate profile generally falls within 
$p\simeq 0.5-0.7$ \citep[e.g.,][]{Yuan_2014, Hu_2022c}, realistic feedback processes are more 
complex and have significant influences on both BH accretion and gas supply from galactic scales. 
Cosmological galaxy-formation simulations indicate that the efficiency of gas removal due to feedback 
is closely linked to the host stellar mass \citep[e.g.,][]{Benson_2010, Hopkins_2012, Somerville_2015, 
Keller_2015,Mayer_2016, Weinberger_2017, Oklopcic_2017, Kim_2019,Qin_2019,Chakraborty_2023}. 
In low-mass galaxies, AGN activity and supernova explosions in their shallow gravitational potential 
efficiently evacuate gas from the nucleus, strongly suppressing BH accretion. In contrast, in 
higher-mass galaxies, the deeper gravitational potential retains gas, reducing the impact of feedback 
and allowing BHs to accrete more efficiently. Given this mass dependence, modeling the evolution of the 
$M_{\rm BH}/M_\star$ ratio under different feedback strengths is crucial for understanding the BH-galaxy 
coevolution.

To capture this mass dependence, we introduce a critical stellar mass at which feedback strength 
varies due to the deepening gravitational potential of galaxies. Motivated by cosmological 
galaxy-formation simulation results, we parameterize the stellar-mass-dependent feedback model 
using a hyperbolic tangent function,
\begin{equation}
    p = p_0-\delta p\cdot \tanh{x},
\end{equation}
where $x = 1.5 \cdot \log_{10}(M_\star/M_{\rm p})$, and $M_{\rm p}$ represents the transition 
mass where feedback strength changes significantly. Due to the complexity of cosmological 
simulations, neither the transition mass nor the feedback strengths for different modes are well 
constrained. We thus adopt the following parameter sets (shown as inserts in Figure~\ref{fig:3}): 
(1) SP1: the transition between strong and moderate feedback occurs at $M_{\rm p}=10^7~\msun$ 
with $p_0=0.75$ and $\delta p=0.25$, 
(2) SP2: the feedback strengths ($p_0$ and $\delta p$) are the same as in SP1, but the transition 
mass is set at $M_{\rm p}=10^9~\msun$, and (3) SP3: the transition occurs at $M_{\rm p}=10^9~\msun$, 
but strong feedback persists to some extent after the transition, with $p_0=0.875$ and $\delta p=0.125$.

In Figure~\ref{fig:3}, we demonstrate that the evolutionary trends of BHs and host galaxies at $z=15$
for these stellar-mass dependent feedback models. The overall evolutionary patterns resemble those 
seen in the constant feedback model in Figure~\ref{fig:2}. Overmassive BHs (relative to the boundary 
line) grow slowly, approaching the boundary, while undermassive BHs grow rapidly toward it. At low 
masses, strong feedback significantly suppresses early BH growth. As the host galaxies evolve, 
weakened feedback allows their central BHs to grow more efficiently, as consistently observed in 
most cosmological simulations. As in the constant feedback models, the evolutionary trajectories 
tend to converge toward the $\mathscr{S}=0$ boundary curves, making them attractors for BH-galaxy 
coevolution. However, in stellar-mass dependent feedback models, the boundary curve shapes differ.
The transition between different feedback strengths at the characteristic mass scale $M_{\rm p}$ 
defines a distinct twisted point on the boundary lines. At higher masses, the final convergent points 
are determined by the feedback strength at $M_\star >M_{p}$. This figure indicates that a weak feedback 
strength as low as $p=0.5$ is sufficient to account for most of the overmassive BHs observed by JWST, 
while stronger feedback ($p\sim 0.7-0.8$) is required to regulate the $M_{\rm BH}/M_\star$ ratios near 
the local relation.

\subsection{How long does seeding information remain?}\label{sec:seed}

Following the evolutionary trends outlined in the previous section, the BH population with 
high $M_{\rm BH}/M_\star$ ratios can be explained by moderate feedback strength during growth.
However, the convergence of these evolutionary trends complicates the interpretation of BH 
seeding models. To address this, we examine the evolutionary tracks for different seeding 
models in Figure~\ref{fig:4}, assuming the SP1 feedback model, and normal distributions for 
$k_{\rm h}$ and $\log\epsilon_{\rm nuc}$. The means and dispersions are set to 
$\langle k_{\rm h}\rangle=0.70$, $\sigma_{k}^2=0.06$, and $\langle \epsilon_{\rm nuc}\rangle=0.02$, 
$\sigma_{\rm nuc}^2=0.25$, respectively (note that the evolutionary pattern is drawn using the 
mean values). We consider two seed models (heavy, and light seeds), whose initial BH and galaxy 
masses at $z=20$ are shown with the shaded regions in the top-left panel of Figure~\ref{fig:4}.
Each seed population evolves according to the evolutionary pattern of the SP1 model. The mass 
distributions at $z=15$, $z=10$, $6$, and $4$ are shown with colored contours, where darker colors 
indicate higher probability densities ($16\%$, and $84\%$). For reference, the JWST-observed 
$M_{\rm BH}-M_\star$ data at each redshift are overlaid as star symbols.

In Figure~\ref{fig:4}, we find that, despite significant differences in seed BH masses, a subset 
of seed BHs from both models grow into the observed overmassive SMBHs at $z \sim 10$. By 
$z \sim 6-4$, all seed BH populations converge toward the boundary line as expected from the 
global evolutionary trend. When the convergence is well established, both seed BH models account 
for the entire distribution of overmassive SMBHs at $z\sim 6$. However, the growth of most heavy 
seed BHs appears stunted, clustering around the characteristic mass scale  $M_{\rm p} = 10^7~M_\odot$. 

Our model suggests that both light and heavy seed BHs can produce overmassive SMBHs through a 
convergent evolutionary path. This implies that the only way to distinguish the origins of seed 
BHs is to trace their early evolution, before they reach convergence $( z \gtrsim 6-10 )$. 
Specifically, detecting BHs with masses of $ \sim 10^{5-6}~\msun $ at higher redshifts that deviate 
from the linear relation set by the convergence phase would provide a key diagnostic, as their 
birth conditions remain imprinted at that stage.

\section{Summary \& Discussions} \label{sec:summary}

In this Letter, we present a BH growth model coupled with DM halo assembly to explore the emergence of 
overmassive BH populations with $M_{\rm BH}/M_\star>0.01$, exceeding the values observed in the local 
universe. We parameterize AGN feedback-driven mass loss as $\dot{M}(r)\propto r^p$ ($0<p<1$), providing 
a semi-analytical framework to map evolutionary trajectories in the $M_{\rm BH}-M_\star$ diagram. The 
growth rates of BHs and galaxies form a converging flow pattern toward a linear attractor set by feedback 
strength $p$ and gas supply efficiency to the nuclei. Overmassive seed BHs grow more slowly than their 
hosts, while undermassive seeds experience rapid growth before aligning with the attractor trajectory.

Our model reproduces the $M_{\rm BH}-M_\star$ distribution of JWST-identified AGNs and BHs at 
$z\sim 4-7$ \citep[e.g.,][]{Kokorev_2023,Bogdan_2024,Greene_2024} across different seeding channels.
However, the convergence behavior erases seeding information from $M_{\rm BH}/M_\star$ ratios 
observed at $z\sim 4-6$. Detecting BHs $\sim 10^{5-6}~\msun$ at higher redshifts that deviate from 
convergence trend would provide key diagnostics of their birth conditions. Meanwhile, we predict 
a broad distribution of $M_{\rm BH}-M_\star$ ratios, some of which might be progenitors for high-$z$ 
submillimeter galaxies (SMGs) and highly obscured AGNs with undermassive SMBHs. Observations on these 
populations will constrain and refine our model parameters.

In addition to the broad distribution, \citet[][]{Zhuang_2023} identified three distinct evolutionary trends 
for local SMBHs—above, on, and below the $M_{\rm BH}-M_\star$ relation. Their analysis suggests that the 
$M_{\rm BH}/M_\star$ ratio in late-type galaxies gradually converges toward this relation, with overmassive 
SMBHs experiencing slow growth and undermassive ones growing more rapidly. This evolutionary pattern aligns 
well with our SP3 model, indicating relatively strong feedback effects in late-type galaxies in the local Universe.

Finally, we caution that despite successfully reproducing the BH-galaxy mass distribution of 
JWST-identified AGNs at high redshifts, our model is based on simplified prescriptions. The uncertainties 
in feedback effects are encapsulated in the power-law form of feedback-driven mass loss, whose physical 
basis derives from numerical simulations and requires further validation through theoretical and 
observational studies. A more detailed model of high-redshift galaxy formation will be essential to 
refine the analytical framework presented in this work.

\begin{acknowledgments}
This work is supported by the Japan Society for the Promotion of Science (JSPS) KAKENHI Grant Number 24KF0130.
We acknowledge support from the National Natural Science Foundation of China (12073003, 12003003, 
11721303, 11991052, 11950410493), and the China Manned Space Project (CMS-CSST-2021-A04 and 
CMS-CSST-2021-A06). L.C.H. is supported by the National Science Foundation of China (12233001), 
the National Key R\&D Program of China (2022YFF0503401). Z.H. acknowledges support by US NSF 
grant AST-2006176 and by NASA grant 80NSSC22K0822. Some of the numerical calculation and analysis 
were performed with the Cray XC50 at the Center for Computational Astrophysics (CfCA) of the 
National Astronomical Observatory of Japan and with the High-performance Computing Platform of 
Peking University.
\end{acknowledgments}

\bibliography{sample631}{}

\begin{thebibliography}{}
\expandafter\ifx\csname natexlab\endcsname\relax\def\natexlab#1{#1}\fi
\providecommand{\url}[1]{\href{#1}{#1}}
\providecommand{\dodoi}[1]{doi:~\href{http://doi.org/#1}{\nolinkurl{#1}}}
\providecommand{\doeprint}[1]{\href{http://ascl.net/#1}{\nolinkurl{http://ascl.net/#1}}}
\providecommand{\doarXiv}[1]{\href{https://arxiv.org/abs/#1}{\nolinkurl{https://arxiv.org/abs/#1}}}

\bibitem[{{Angl{\'e}s-Alc{\'a}zar} {et~al.}(2017){Angl{\'e}s-Alc{\'a}zar},
  {Faucher-Gigu{\`e}re}, {Quataert}, {Hopkins}, {Feldmann}, {Torrey}, {Wetzel},
  \& {Kere{\v{s}}}}]{Angles-Alcazar_2017}
{Angl{\'e}s-Alc{\'a}zar}, D., {Faucher-Gigu{\`e}re}, C.-A., {Quataert}, E.,
  {et~al.} 2017, \mnras, 472, L109, \dodoi{10.1093/mnrasl/slx161}

\bibitem[{Ba{\~n}ados {et~al.}(2018)Ba{\~n}ados, Venemans, Mazzucchelli,
  {et~al.}}]{Banados_2018}
Ba{\~n}ados, E., Venemans, B.~P., Mazzucchelli, C., {et~al.} 2018, Nature, 553,
  473, \dodoi{10.1038/nature25180}

\bibitem[{{Behroozi} {et~al.}(2019){Behroozi}, {Wechsler}, {Hearin}, \&
  {Conroy}}]{Behroozi_2019}
{Behroozi}, P., {Wechsler}, R.~H., {Hearin}, A.~P., \& {Conroy}, C. 2019,
  Monthly Notices of the Royal Astronomical Society, 488, 3143,
  \dodoi{10.1093/mnras/stz1182}

\bibitem[{{Benson}(2010)}]{Benson_2010}
{Benson}, A.~J. 2010, \physrep, 495, 33, \dodoi{10.1016/j.physrep.2010.06.001}

\bibitem[{{Bogd{\'a}n} {et~al.}(2024){Bogd{\'a}n}, {Goulding}, {Natarajan},
  {Kov{\'a}cs}, {Tremblay}, {Chadayammuri}, {Volonteri}, {Kraft}, {Forman},
  {Jones}, {Churazov}, \& {Zhuravleva}}]{Bogdan_2024}
{Bogd{\'a}n}, {\'A}., {Goulding}, A.~D., {Natarajan}, P., {et~al.} 2024, Nature
  Astronomy, 8, 126, \dodoi{10.1038/s41550-023-02111-9}

\bibitem[{{Bromm} \& {Loeb}(2003)}]{Bromm_2003}
{Bromm}, V., \& {Loeb}, A. 2003, \apj, 596, 34, \dodoi{10.1086/377529}

\bibitem[{{Chakraborty} {et~al.}(2023){Chakraborty}, {Chatterjee}, {Lacy},
  {Roy}, {Roy}, \& {Kar Chowdhury}}]{Chakraborty_2023}
{Chakraborty}, A., {Chatterjee}, S., {Lacy}, M., {et~al.} 2023, \apj, 954, 8,
  \dodoi{10.3847/1538-4357/ace1e4}

\bibitem[{{Chen} {et~al.}(2024){Chen}, {Ho}, {Li}, \& {Zhuang}}]{Chen_2024}
{Chen}, C.-H., {Ho}, L.~C., {Li}, R., \& {Zhuang}, M.-Y. 2024, arXiv e-prints,
  arXiv:2411.04446, \dodoi{10.48550/arXiv.2411.04446}

\bibitem[{{Cole} {et~al.}(2000){Cole}, {Lacey}, {Baugh}, \&
  {Frenk}}]{Cole_2000}
{Cole}, S., {Lacey}, C.~G., {Baugh}, C.~M., \& {Frenk}, C.~S. 2000, \mnras,
  319, 168, \dodoi{10.1046/j.1365-8711.2000.03879.x}

\bibitem[{{Dekel} {et~al.}(2013){Dekel}, {Zolotov}, {Tweed}, {Cacciato},
  {Ceverino}, \& {Primack}}]{Dekel_2013}
{Dekel}, A., {Zolotov}, A., {Tweed}, D., {et~al.} 2013, \mnras, 435, 999,
  \dodoi{10.1093/mnras/stt1338}

\bibitem[{{Ding} {et~al.}(2023){Ding}, {Onoue}, {Silverman}, {Matsuoka},
  {Izumi}, {Strauss}, {Jahnke}, {Phillips}, {Li}, {Volonteri}, {Haiman},
  {Andika}, {Aoki}, {Baba}, {Bieri}, {Bosman}, {Bottrell}, {Eilers},
  {Fujimoto}, {Habouzit}, {Imanishi}, {Inayoshi}, {Iwasawa}, {Kashikawa},
  {Kawaguchi}, {Kohno}, {Lee}, {Lupi}, {Lyu}, {Nagao}, {Overzier}, {Schindler},
  {Schramm}, {Shimasaku}, {Toba}, {Trakhtenbrot}, {Trebitsch}, {Treu},
  {Umehata}, {Venemans}, {Vestergaard}, {Walter}, {Wang}, \&
  {Yang}}]{Ding_2023}
{Ding}, X., {Onoue}, M., {Silverman}, J.~D., {et~al.} 2023, \nat, 621, 51,
  \dodoi{10.1038/s41586-023-06345-5}

\bibitem[{{Dubois} {et~al.}(2015){Dubois}, {Volonteri}, {Silk}, {Devriendt},
  {Slyz}, \& {Teyssier}}]{Dubois_2015}
{Dubois}, Y., {Volonteri}, M., {Silk}, J., {et~al.} 2015, \mnras, 452, 1502,
  \dodoi{10.1093/mnras/stv1416}

\bibitem[{{Fakhouri} {et~al.}(2010){Fakhouri}, {Ma}, \&
  {Boylan-Kolchin}}]{Fakhouri_2010}
{Fakhouri}, O., {Ma}, C.-P., \& {Boylan-Kolchin}, M. 2010, \mnras, 406, 2267,
  \dodoi{10.1111/j.1365-2966.2010.16859.x}

\bibitem[{Fan {et~al.}(2023)Fan, Ba{\~n}ados, \& Simcoe}]{Fan_2023}
Fan, X., Ba{\~n}ados, E., \& Simcoe, R.~A. 2023, Annual Review of Astronomy and
  Astrophysics, 61, 373

\bibitem[{{Fraternali} \& {Binney}(2006)}]{Fraternali_2006}
{Fraternali}, F., \& {Binney}, J.~J. 2006, \mnras, 366, 449,
  \dodoi{10.1111/j.1365-2966.2005.09816.x}

\bibitem[{{Fraternali} {et~al.}(2005){Fraternali}, {Oosterloo}, {Sancisi}, \&
  {Swaters}}]{Fraternali_2005}
{Fraternali}, F., {Oosterloo}, T.~A., {Sancisi}, R., \& {Swaters}, R. 2005, in
  Astronomical Society of the Pacific Conference Series, Vol. 331, Extra-Planar
  Gas, ed. R.~{Braun}, 239, \dodoi{10.48550/arXiv.astro-ph/0410375}

\bibitem[{{Greene} {et~al.}(2020){Greene}, {Strader}, \& {Ho}}]{Greene_2020}
{Greene}, J.~E., {Strader}, J., \& {Ho}, L.~C. 2020, \araa, 58, 257,
  \dodoi{10.1146/annurev-astro-032620-021835}

\bibitem[{{Greene} {et~al.}(2024){Greene}, {Labbe}, {Goulding}, {Furtak},
  {Chemerynska}, {Kokorev}, {Dayal}, {Volonteri}, {Williams}, {Wang}, {Setton},
  {Burgasser}, {Bezanson}, {Atek}, {Brammer}, {Cutler}, {Feldmann}, {Fujimoto},
  {Glazebrook}, {de Graaff}, {Khullar}, {Leja}, {Marchesini}, {Maseda},
  {Matthee}, {Miller}, {Naidu}, {Nanayakkara}, {Oesch}, {Pan}, {Papovich},
  {Price}, {van Dokkum}, {Weaver}, {Whitaker}, \& {Zitrin}}]{Greene_2024}
{Greene}, J.~E., {Labbe}, I., {Goulding}, A.~D., {et~al.} 2024, \apj, 964, 39,
  \dodoi{10.3847/1538-4357/ad1e5f}

\bibitem[{{Habouzit}(2025)}]{Habouzit_2025}
{Habouzit}, M. 2025, \mnras, \dodoi{10.1093/mnras/staf167}

\bibitem[{{Habouzit} {et~al.}(2017){Habouzit}, {Volonteri}, \&
  {Dubois}}]{Habouzit_2017}
{Habouzit}, M., {Volonteri}, M., \& {Dubois}, Y. 2017, \mnras, 468, 3935,
  \dodoi{10.1093/mnras/stx666}

\bibitem[{{Haiman} \& {Loeb}(2001)}]{Haiman_2001}
{Haiman}, Z., \& {Loeb}, A. 2001, \apj, 552, 459, \dodoi{10.1086/320586}

\bibitem[{{Harikane} {et~al.}(2023){Harikane}, {Zhang}, {Nakajima}, {Ouchi},
  {Isobe}, {Ono}, {Hatano}, {Xu}, \& {Umeda}}]{Harikane_2023}
{Harikane}, Y., {Zhang}, Y., {Nakajima}, K., {et~al.} 2023, \apj, 959, 39,
  \dodoi{10.3847/1538-4357/ad029e}

\bibitem[{Hopkins \& Quataert(2010)}]{Hopkins_2010}
Hopkins, P.~F., \& Quataert, E. 2010, Monthly Notices of the Royal Astronomical
  Society, 407, 1529, \dodoi{10.1111/j.1365-2966.2010.17064.x}

\bibitem[{{Hopkins} {et~al.}(2012){Hopkins}, {Quataert}, \&
  {Murray}}]{Hopkins_2012}
{Hopkins}, P.~F., {Quataert}, E., \& {Murray}, N. 2012, \mnras, 421, 3522,
  \dodoi{10.1111/j.1365-2966.2012.20593.x}

\bibitem[{Hu {et~al.}(2022)Hu, Inayoshi, Haiman, Li, Quataert, \&
  Kuiper}]{Hu_2022c}
Hu, H., Inayoshi, K., Haiman, Z., {et~al.} 2022, The Astrophysical Journal,
  935, 140, \dodoi{10.3847/1538-4357/ac7daa}

\bibitem[{Inayoshi {et~al.}(2022)Inayoshi, Harikane, Inoue, Li, \&
  Ho}]{Inayoshi_2022b}
Inayoshi, K., Harikane, Y., Inoue, A.~K., Li, W., \& Ho, L.~C. 2022, The
  Astrophysical Journal Letters, 938, L10, \dodoi{10.3847/2041-8213/ac9310}

\bibitem[{{Inayoshi} {et~al.}(2022){Inayoshi}, {Nakatani}, {Toyouchi},
  {Hosokawa}, {Kuiper}, \& {Onoue}}]{Inayoshi_2022a}
{Inayoshi}, K., {Nakatani}, R., {Toyouchi}, D., {et~al.} 2022, The
  Astrophysical Journal, 927, 237, \dodoi{10.3847/1538-4357/ac4751}

\bibitem[{Inayoshi {et~al.}(2020)Inayoshi, Visbal, \& Haiman}]{Inayoshi_2020}
Inayoshi, K., Visbal, E., \& Haiman, Z. 2020, Annual Review of Astronomy and
  Astrophysics, 58, 27, \dodoi{10.1146/annurev-astro-120419-014455}

\bibitem[{{Izumi} {et~al.}(2019){Izumi}, {Onoue}, {Matsuoka}, {Nagao},
  {Strauss}, {Imanishi}, {Kashikawa}, {Fujimoto}, {Kohno}, {Toba}, {Umehata},
  {Goto}, {Ueda}, {Shirakata}, {Silverman}, {Greene}, {Harikane}, {Hashimoto},
  {Ikarashi}, {Iono}, {Iwasawa}, {Lee}, {Minezaki}, {Nakanishi}, {Tamura},
  {Tang}, \& {Taniguchi}}]{Izumi_2019}
{Izumi}, T., {Onoue}, M., {Matsuoka}, Y., {et~al.} 2019, \pasj, 71, 111,
  \dodoi{10.1093/pasj/psz096}

\bibitem[{Juodžbalis {et~al.}(2024)Juodžbalis, Maiolino, Baker, Tacchella,
  Scholtz, D'Eugenio, Schneider, Trinca, Valiante, DeCoursey, Curti, Carniani,
  Chevallard, de~Graaff, Arribas, Bennett, Bourne, Bunker, Charlot, Jiang,
  Koudmani, Perna, Robertson, Sijacki, Übler, Williams, Willott, \&
  Witstok}]{Juodzbalis_2024}
Juodžbalis, I., Maiolino, R., Baker, W.~M., {et~al.} 2024, A dormant,
  overmassive black hole in the early {Universe},  arXiv.
\newblock \url{http://arxiv.org/abs/2403.03872}

\bibitem[{{Kaufmann} {et~al.}(2006){Kaufmann}, {Mayer}, {Wadsley}, {Stadel}, \&
  {Moore}}]{Kaufmann_2006}
{Kaufmann}, T., {Mayer}, L., {Wadsley}, J., {Stadel}, J., \& {Moore}, B. 2006,
  \mnras, 370, 1612, \dodoi{10.1111/j.1365-2966.2006.10599.x}

\bibitem[{{Keller} {et~al.}(2015){Keller}, {Wadsley}, \&
  {Couchman}}]{Keller_2015}
{Keller}, B.~W., {Wadsley}, J., \& {Couchman}, H.~M.~P. 2015, \mnras, 453,
  3499, \dodoi{10.1093/mnras/stv1789}

\bibitem[{{Kim} {et~al.}(2019){Kim}, {Wise}, {Abel}, {Jo}, {Primack}, \&
  {Hopkins}}]{Kim_2019}
{Kim}, J.-h., {Wise}, J.~H., {Abel}, T., {et~al.} 2019, \apj, 887, 120,
  \dodoi{10.3847/1538-4357/ab510b}

\bibitem[{{Kocevski} {et~al.}(2023){Kocevski}, {Onoue}, {Inayoshi}, {Trump},
  {Arrabal Haro}, {Grazian}, {Dickinson}, {Finkelstein}, {Kartaltepe},
  {Hirschmann}, {Aird}, {Holwerda}, {Fujimoto}, {Juneau}, {Amor{\'\i}n},
  {Backhaus}, {Bagley}, {Barro}, {Bell}, {Bisigello}, {Calabr{\`o}}, {Cleri},
  {Cooper}, {Ding}, {Grogin}, {Ho}, {Hutchison}, {Inoue}, {Jiang}, {Jones},
  {Koekemoer}, {Li}, {Li}, {McGrath}, {Molina}, {Papovich},
  {P{\'e}rez-Gonz{\'a}lez}, {Pirzkal}, {Wilkins}, {Yang}, \&
  {Yung}}]{Kocevski_2023}
{Kocevski}, D.~D., {Onoue}, M., {Inayoshi}, K., {et~al.} 2023, \apjl, 954, L4,
  \dodoi{10.3847/2041-8213/ace5a0}

\bibitem[{{Kocevski} {et~al.}(2024){Kocevski}, {Finkelstein}, {Barro},
  {Taylor}, {Calabr{\`o}}, {Laloux}, {Buchner}, {Trump}, {Leung}, {Yang},
  {Dickinson}, {P{\'e}rez-Gonz{\'a}lez}, {Pacucci}, {Inayoshi}, {Somerville},
  {McGrath}, {Akins}, {Bagley}, {Bisigello}, {Bowler}, {Carnall}, {Casey},
  {Cheng}, {Cleri}, {Costantin}, {Cullen}, {Davis}, {Donnan}, {Dunlop},
  {Ellis}, {Ferguson}, {Fujimoto}, {Fontana}, {Giavalisco}, {Grazian},
  {Grogin}, {Hathi}, {Hirschmann}, {Huertas-Company}, {Holwerda},
  {Illingworth}, {Juneau}, {Kartaltepe}, {Koekemoer}, {Li}, {Lucas}, {Magee},
  {Mason}, {McLeod}, {McLure}, {Napolitano}, {Papovich}, {Pirzkal},
  {Rodighiero}, {Santini}, {Wilkins}, \& {Yung}}]{Kocevski_2024}
{Kocevski}, D.~D., {Finkelstein}, S.~L., {Barro}, G., {et~al.} 2024, arXiv
  e-prints, arXiv:2404.03576, \dodoi{10.48550/arXiv.2404.03576}

\bibitem[{{Kokorev} {et~al.}(2023){Kokorev}, {Fujimoto}, {Labbe}, {Greene},
  {Bezanson}, {Dayal}, {Nelson}, {Atek}, {Brammer}, {Caputi}, {Chemerynska},
  {Cutler}, {Feldmann}, {Fudamoto}, {Furtak}, {Goulding}, {de Graaff}, {Leja},
  {Marchesini}, {Miller}, {Nanayakkara}, {Oesch}, {Pan}, {Price}, {Setton},
  {Smit}, {Stefanon}, {Wang}, {Weaver}, {Whitaker}, {Williams}, \&
  {Zitrin}}]{Kokorev_2023}
{Kokorev}, V., {Fujimoto}, S., {Labbe}, I., {et~al.} 2023, \apjl, 957, L7,
  \dodoi{10.3847/2041-8213/ad037a}

\bibitem[{Kormendy \& Ho(2013)}]{Kormendy_2013}
Kormendy, J., \& Ho, L.~C. 2013, Annual Review of Astronomy and Astrophysics,
  51, 511, \dodoi{10.1146/annurev-astro-082708-101811}

\bibitem[{{Lacey} \& {Cole}(1993)}]{Lacey_1993}
{Lacey}, C., \& {Cole}, S. 1993, \mnras, 262, 627,
  \dodoi{10.1093/mnras/262.3.627}

\bibitem[{{Larson} {et~al.}(2023){Larson}, {Finkelstein}, {Kocevski},
  {Hutchison}, {Trump}, {Arrabal Haro}, {Bromm}, {Cleri}, {Dickinson},
  {Fujimoto}, {Kartaltepe}, {Koekemoer}, {Papovich}, {Pirzkal}, {Tacchella},
  {Zavala}, {Bagley}, {Behroozi}, {Champagne}, {Cole}, {Jung}, {Morales},
  {Yang}, {Zhang}, {Zitrin}, {Amor{\'\i}n}, {Burgarella}, {Casey}, {Ch{\'a}vez
  Ortiz}, {Cox}, {Chworowsky}, {Fontana}, {Gawiser}, {Grazian}, {Grogin},
  {Harish}, {Hathi}, {Hirschmann}, {Holwerda}, {Juneau}, {Leung}, {Lucas},
  {McGrath}, {P{\'e}rez-Gonz{\'a}lez}, {Rigby}, {Seill{\'e}}, {Simons}, {de La
  Vega}, {Weiner}, {Wilkins}, {Yung}, \& {Ceers Team}}]{Larson_2023}
{Larson}, R.~L., {Finkelstein}, S.~L., {Kocevski}, D.~D., {et~al.} 2023, \apjl,
  953, L29, \dodoi{10.3847/2041-8213/ace619}

\bibitem[{{Li} {et~al.}(2024{\natexlab{a}}){Li}, {Silverman}, {Shen},
  {Volonteri}, {Jahnke}, {Zhuang}, {Scoggins}, {Ding}, {Harikane}, {Onoue}, \&
  {Tanaka}}]{Li_2024}
{Li}, J., {Silverman}, J.~D., {Shen}, Y., {et~al.} 2024{\natexlab{a}}, arXiv
  e-prints, arXiv:2403.00074, \dodoi{10.48550/arXiv.2403.00074}

\bibitem[{{Li} {et~al.}(2023){Li}, {Inayoshi}, {Onoue}, \&
  {Toyouchi}}]{Li_2023a}
{Li}, W., {Inayoshi}, K., {Onoue}, M., \& {Toyouchi}, D. 2023, \apj, 950, 85,
  \dodoi{10.3847/1538-4357/accbbe}

\bibitem[{Li {et~al.}(2021)Li, Inayoshi, \& Qiu}]{Li_2021}
Li, W., Inayoshi, K., \& Qiu, Y. 2021, The Astrophysical Journal, 917, 60,
  \dodoi{10.3847/1538-4357/ac0adc}

\bibitem[{{Li} {et~al.}(2024{\natexlab{b}}){Li}, {Inayoshi}, {Onoue}, {He},
  {Matsuoka}, {Pan}, {Akiyama}, {Izumi}, \& {Nagao}}]{Li_2023}
{Li}, W., {Inayoshi}, K., {Onoue}, M., {et~al.} 2024{\natexlab{b}}, \apj, 969,
  69, \dodoi{10.3847/1538-4357/ad46f9}

\bibitem[{{Madau} \& {Rees}(2001)}]{Madau_2001}
{Madau}, P., \& {Rees}, M.~J. 2001, \apjl, 551, L27, \dodoi{10.1086/319848}

\bibitem[{Maiolino {et~al.}(2023)Maiolino, Scholtz, Curtis-Lake,
  {et~al.}}]{Maiolino_2023}
Maiolino, R., Scholtz, J., Curtis-Lake, E., {et~al.} 2023, {JADES}. {The}
  diverse population of infant {Black} {Holes} at 4{\textless}z{\textless}11:
  merging, tiny, poor, but mighty,  arXiv.
\newblock \url{http://arxiv.org/abs/2308.01230}

\bibitem[{{Maiolino} {et~al.}(2024){Maiolino}, {Scholtz}, {Witstok},
  {Carniani}, {D'Eugenio}, {de Graaff}, {{\"U}bler}, {Tacchella},
  {Curtis-Lake}, {Arribas}, {Bunker}, {Charlot}, {Chevallard}, {Curti},
  {Looser}, {Maseda}, {Rawle}, {Rodr{\'\i}guez del Pino}, {Willott}, {Egami},
  {Eisenstein}, {Hainline}, {Robertson}, {Williams}, {Willmer}, {Baker},
  {Boyett}, {DeCoursey}, {Fabian}, {Helton}, {Ji}, {Jones}, {Kumari},
  {Laporte}, {Nelson}, {Perna}, {Sandles}, {Shivaei}, \& {Sun}}]{Maiolino_2024}
{Maiolino}, R., {Scholtz}, J., {Witstok}, J., {et~al.} 2024, \nat, 627, 59,
  \dodoi{10.1038/s41586-024-07052-5}

\bibitem[{Mayer {et~al.}(2016)Mayer, Tamburello, Lupi, Keller, Wadsley, \&
  Madau}]{Mayer_2016}
Mayer, L., Tamburello, V., Lupi, A., {et~al.} 2016, The Astrophysical Journal
  Letters, 830, L13

\bibitem[{Mortlock {et~al.}(2011)Mortlock, Warren, Venemans,
  {et~al.}}]{Mortlock_2011}
Mortlock, D.~J., Warren, S.~J., Venemans, B.~P., {et~al.} 2011, Nature, 474,
  616, \dodoi{10.1038/nature10159}

\bibitem[{{Oklop{\v{c}}i{\'c}} {et~al.}(2017){Oklop{\v{c}}i{\'c}}, {Hopkins},
  {Feldmann}, {Kere{\v{s}}}, {Faucher-Gigu{\`e}re}, \&
  {Murray}}]{Oklopcic_2017}
{Oklop{\v{c}}i{\'c}}, A., {Hopkins}, P.~F., {Feldmann}, R., {et~al.} 2017,
  \mnras, 465, 952, \dodoi{10.1093/mnras/stw2754}

\bibitem[{{Onoue} {et~al.}(2023){Onoue}, {Inayoshi}, {Ding}, {Li}, {Li},
  {Molina}, {Inoue}, {Jiang}, \& {Ho}}]{Onoue_2023}
{Onoue}, M., {Inayoshi}, K., {Ding}, X., {et~al.} 2023, \apjl, 942, L17,
  \dodoi{10.3847/2041-8213/aca9d3}

\bibitem[{Parkinson {et~al.}(2008)Parkinson, Cole, \& Helly}]{Parkinson_2008}
Parkinson, H., Cole, S., \& Helly, J. 2008, Monthly Notices of the Royal
  Astronomical Society, 383, 557

\bibitem[{{Planck Collaboration} {et~al.}(2020){Planck Collaboration}, Aghanim,
  Akrami, Ashdown, Aumont, Baccigalupi, Ballardini, Banday, Barreiro, Bartolo,
  Basak, Battye, Benabed, Bernard, Bersanelli, Bielewicz, Bock, Bond, Borrill,
  Bouchet, Boulanger, Bucher, Burigana, Butler, Calabrese, Cardoso, Carron,
  Challinor, Chiang, Chluba, Colombo, Combet, Contreras, Crill, Cuttaia,
  De~Bernardis, De~Zotti, Delabrouille, Delouis, Di~Valentino, Diego, Doré,
  Douspis, Ducout, Dupac, Dusini, Efstathiou, Elsner, Enßlin, Eriksen,
  Fantaye, Farhang, Fergusson, Fernandez-Cobos, Finelli, Forastieri, Frailis,
  Fraisse, Franceschi, Frolov, Galeotta, Galli, Ganga, Génova-Santos, Gerbino,
  Ghosh, González-Nuevo, Górski, Gratton, Gruppuso, Gudmundsson, Hamann,
  Handley, Hansen, Herranz, Hildebrandt, Hivon, Huang, Jaffe, Jones, Karakci,
  Keihänen, Keskitalo, Kiiveri, Kim, Kisner, Knox, Krachmalnicoff, Kunz,
  Kurki-Suonio, Lagache, Lamarre, Lasenby, Lattanzi, Lawrence, Le~Jeune, Lemos,
  Lesgourgues, Levrier, Lewis, Liguori, Lilje, Lilley, Lindholm,
  López-Caniego, Lubin, Ma, Macías-Pérez, Maggio, Maino, Mandolesi,
  Mangilli, Marcos-Caballero, Maris, Martin, Martinelli, Martínez-González,
  Matarrese, Mauri, McEwen, Meinhold, Melchiorri, Mennella, Migliaccio, Millea,
  Mitra, Miville-Deschênes, Molinari, Montier, Morgante, Moss, Natoli,
  Nørgaard-Nielsen, Pagano, Paoletti, Partridge, Patanchon, Peiris, Perrotta,
  Pettorino, Piacentini, Polastri, Polenta, Puget, Rachen, Reinecke,
  Remazeilles, Renzi, Rocha, Rosset, Roudier, Rubiño-Martín, Ruiz-Granados,
  Salvati, Sandri, Savelainen, Scott, Shellard, Sirignano, Sirri, Spencer,
  Sunyaev, Suur-Uski, Tauber, Tavagnacco, Tenti, Toffolatti, Tomasi, Trombetti,
  Valenziano, Valiviita, Van~Tent, Vibert, Vielva, Villa, Vittorio, Wandelt,
  Wehus, White, White, Zacchei, \& Zonca}]{Planck_2020}
{Planck Collaboration}, Aghanim, N., Akrami, Y., {et~al.} 2020, Astronomy \&
  Astrophysics, 641, A6, \dodoi{10.1051/0004-6361/201833910}

\bibitem[{{Pontoppidan} {et~al.}(2022){Pontoppidan}, {Barrientes}, {Blome},
  {Braun}, {Brown}, {Carruthers}, {Coe}, {DePasquale}, {Espinoza}, {Marin},
  {Gordon}, {Henry}, {Hustak}, {James}, {Jenkins}, {Koekemoer}, {LaMassa},
  {Law}, {Lockwood}, {Moro-Martin}, {Mullally}, {Pagan}, {Player}, {Proffitt},
  {Pulliam}, {Ramsay}, {Ravindranath}, {Reid}, {Robberto}, {Sabbi}, {Ubeda},
  {Balogh}, {Flanagan}, {Gardner}, {Hasan}, {Meinke}, \&
  {Nota}}]{Pontoppidan_2022}
{Pontoppidan}, K.~M., {Barrientes}, J., {Blome}, C., {et~al.} 2022, \apjl, 936,
  L14, \dodoi{10.3847/2041-8213/ac8a4e}

\bibitem[{{Press} \& {Schechter}(1974)}]{Press_1974}
{Press}, W.~H., \& {Schechter}, P. 1974, \apj, 187, 425, \dodoi{10.1086/152650}

\bibitem[{{Qin} {et~al.}(2019){Qin}, {Duffy}, {Mutch}, {Poole}, {Mesinger}, \&
  {Wyithe}}]{Qin_2019}
{Qin}, Y., {Duffy}, A.~R., {Mutch}, S.~J., {et~al.} 2019, \mnras, 487, 1946,
  \dodoi{10.1093/mnras/stz1380}

\bibitem[{{Regan} \& {Haehnelt}(2009{\natexlab{a}})}]{Regan_2009a}
{Regan}, J.~A., \& {Haehnelt}, M.~G. 2009{\natexlab{a}}, \mnras, 396, 343,
  \dodoi{10.1111/j.1365-2966.2009.14579.x}

\bibitem[{{Regan} \& {Haehnelt}(2009{\natexlab{b}})}]{Regan_2009b}
---. 2009{\natexlab{b}}, \mnras, 393, 858,
  \dodoi{10.1111/j.1365-2966.2008.14088.x}

\bibitem[{{Reines} \& {Volonteri}(2015)}]{Reines_2015}
{Reines}, A.~E., \& {Volonteri}, M. 2015, The Astrophysical Journal, 813, 82,
  \dodoi{10.1088/0004-637X/813/2/82}

\bibitem[{{Scoggins} \& {Haiman}(2024)}]{Scoggins_2024}
{Scoggins}, M.~T., \& {Haiman}, Z. 2024, \mnras, 531, 4584,
  \dodoi{10.1093/mnras/stae1449}

\bibitem[{{Somerville} \& {Dav{\'e}}(2015)}]{Somerville_2015}
{Somerville}, R.~S., \& {Dav{\'e}}, R. 2015, \araa, 53, 51,
  \dodoi{10.1146/annurev-astro-082812-140951}

\bibitem[{{Stone} {et~al.}(2024){Stone}, {Lyu}, {Rieke}, {Alberts}, \&
  {Hainline}}]{Stone_2024}
{Stone}, M.~A., {Lyu}, J., {Rieke}, G.~H., {Alberts}, S., \& {Hainline}, K.~N.
  2024, \apj, 964, 90, \dodoi{10.3847/1538-4357/ad2a57}

\bibitem[{{Trinca} {et~al.}(2022){Trinca}, {Schneider}, {Valiante}, {Graziani},
  {Zappacosta}, \& {Shankar}}]{Trinca_2022}
{Trinca}, A., {Schneider}, R., {Valiante}, R., {et~al.} 2022, \mnras, 511, 616,
  \dodoi{10.1093/mnras/stac062}

\bibitem[{{van den Bosch} {et~al.}(2012){van den Bosch}, {Gebhardt},
  {G{\"u}ltekin}, {van de Ven}, {van der Wel}, \& {Walsh}}]{van_2012}
{van den Bosch}, R. C.~E., {Gebhardt}, K., {G{\"u}ltekin}, K., {et~al.} 2012,
  \nat, 491, 729, \dodoi{10.1038/nature11592}

\bibitem[{{Volonteri} {et~al.}(2003){Volonteri}, {Haardt}, \&
  {Madau}}]{Volonteri_2003}
{Volonteri}, M., {Haardt}, F., \& {Madau}, P. 2003, \apj, 582, 559,
  \dodoi{10.1086/344675}

\bibitem[{{Volonteri} {et~al.}(2021){Volonteri}, {Habouzit}, \&
  {Colpi}}]{Volonteri_2021}
{Volonteri}, M., {Habouzit}, M., \& {Colpi}, M. 2021, Nature Reviews Physics,
  3, 732, \dodoi{10.1038/s42254-021-00364-9}

\bibitem[{{Wadsley} {et~al.}(2004){Wadsley}, {Stadel}, \&
  {Quinn}}]{Wadsley_2004}
{Wadsley}, J.~W., {Stadel}, J., \& {Quinn}, T. 2004, \na, 9, 137,
  \dodoi{10.1016/j.newast.2003.08.004}

\bibitem[{{Wang} {et~al.}(2023){Wang}, {Lei}, {Yuan}, \& {Fan}}]{Wang_2023}
{Wang}, Y.-Y., {Lei}, L., {Yuan}, G.-W., \& {Fan}, Y.-Z. 2023, \apjl, 954, L48,
  \dodoi{10.3847/2041-8213/acf46c}

\bibitem[{{Weinberger} {et~al.}(2017){Weinberger}, {Springel}, {Hernquist},
  {Pillepich}, {Marinacci}, {Pakmor}, {Nelson}, {Genel}, {Vogelsberger},
  {Naiman}, \& {Torrey}}]{Weinberger_2017}
{Weinberger}, R., {Springel}, V., {Hernquist}, L., {et~al.} 2017, \mnras, 465,
  3291, \dodoi{10.1093/mnras/stw2944}

\bibitem[{Wu {et~al.}(2015)Wu, Wang, Fan, {et~al.}}]{Wu_2015}
Wu, X.-B., Wang, F., Fan, X., {et~al.} 2015, Nature, 518, 512,
  \dodoi{10.1038/nature14241}

\bibitem[{{Wyithe} \& {Padmanabhan}(2006)}]{Wyithe_2006}
{Wyithe}, J. S.~B., \& {Padmanabhan}, T. 2006, \mnras, 366, 1029,
  \dodoi{10.1111/j.1365-2966.2005.09858.x}

\bibitem[{{Yuan} \& {Narayan}(2014)}]{Yuan_2014}
{Yuan}, F., \& {Narayan}, R. 2014, \araa, 52, 529,
  \dodoi{10.1146/annurev-astro-082812-141003}

\bibitem[{Yue {et~al.}(2023)Yue, Eilers, Simcoe, Mackenzie, Matthee, Kashino,
  Bordoloi, Lilly, \& Naidu}]{Yue_2023}
Yue, M., Eilers, A.-C., Simcoe, R.~A., {et~al.} 2023, {EIGER} {V}.
  {Characterizing} the {Host} {Galaxies} of {Luminous} {Quasars} at
  \$z{\textbackslash}gtrsim6\$,  arXiv.
\newblock \url{http://arxiv.org/abs/2309.04614}

\bibitem[{{Zhuang} \& {Ho}(2023)}]{Zhuang_2023}
{Zhuang}, M.-Y., \& {Ho}, L.~C. 2023, Nature Astronomy, 7, 1376,
  \dodoi{10.1038/s41550-023-02051-4}

\end{thebibliography}
\bibliographystyle{aasjournal}

\end{CJK*}
\end{document}